\documentclass[letterpaper,12pt]{article}   
\usepackage{osajnl2} 
\usepackage[draft]{hyperref} 
\usepackage{amsmath}
\usepackage{graphicx}
\usepackage{epstopdf}

\begin{document}

\title{Layer-oriented adaptive optics for solar telescopes}

\author{Agla\'e Kellerer}
\address{Big Bear Solar Observatory, \\ 
40 386 North Shore Lane, Big Bear City, CA 92314 - 9672, USA\\
kellerer@bbso.njit.edu}


\begin{abstract} 

First multi-conjugate adaptive-optical (MCAO) systems are currently being installed on solar telescopes. The aim of these systems is to increase the corrected field-of-view with respect to conventional adaptive optics. 
However, this first generation is based on a star-oriented approach, and it is then difficult to increase the size of the field-of-view beyond $60''-80''$ in diameter. 
We propose to implement the layer-oriented approach in solar MCAO by use of wide-field Shack-Hartmann wavefront sensors conjugated to the strongest turbulent layers. 
The wavefront distortions are averaged over a wide-field: the signal from distant turbulence is attenuated and the tomographic reconstruction is thus done optically.  
The system consists of independent correction loops, that only need to account for local turbulence: the sub-apertures can be enlarged and the correction frequency reduced.
Most importantly, a star-oriented MCAO system becomes more complex with increasing field size, while the layer-oriented approach benefits from larger fields -- and will therefore be an attractive solution for the future generation of solar MCAO systems. 
\end{abstract}


\maketitle 

\section{Introduction}
\subsection{The current approach to solar MCAO}

The need for exact solar observations has led to the development of adaptive-optical (AO) correction systems on solar telescopes\,\cite{Rimmele2011}.
Classical AO systems are limited by the isoplanatic angle of atmospheric turbulence and solar images are only corrected within, typically,  10\,arcseconds. 
In multi-conjugate adaptive optical (MCAO) systems, the use of several deformable mirrors allows to correct the wavefront distortions within larger field-of-views. Each mirror is optically conjugated to a different altitude and a phase-distortion produced at height $h$ is then corrected by the mirror whose {\it conjugate altitude\/} is closest to $h$. This ideal scenario is in practice limited by our incomplete knowledge of the altitude distribution of turbulence. In this article we suggest to improve the sensing stage of solar MCAO systems by implementing a {\it layer-oriented approach\/}\,\cite{Ragazzoni}, \cite{Ragazzoni2}, \cite{Arcid}.

\begin{figure}[htbp]
\begin{center}
\includegraphics[width=.6\textwidth]{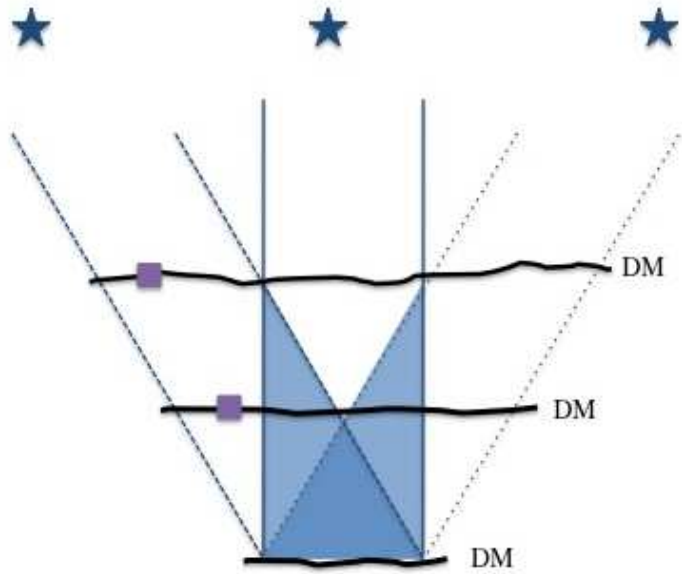}
\caption{ Problematic of the star-oriented approach. Outside the blue, M-shaped area, there are no redundant measurements: Turbulence in the two purple squares, for example, yield the same sensor measurement and can not be distinguished. The solution consists in applying a correction that will -- on average -- be more right than wrong: the control-loop is informed about the Kolmogorov nature of turbulence and about the current profile of atmospheric turbulence. DM: deformable mirror. }
\label{fig:triang}
\end{center}
\end{figure}

Until now, solar MCAO experiments have exclusively used a {\it star-oriented approach\/} where each sensor measures wavefront distortions within a narrow field (on a star in nighttime astronomy, hence the name). The height distribution of turbulence is then determined via tomography. 
This approach has been demonstrated successfully on several telescopes, e.g. the VTT on the Canary Islands\,\cite{LŸhe2005} and the DST in New Mexico\,\cite{Rimmele2010}. The atmospheric turbulence was corrected within roughly $45''\times 45''$ compared to $10''\times 10''$ for traditional AO systems.
However, in the star-oriented approach larger field-sizes will be difficult to correct. Berkefeld et al.\,\cite{Berkefeld} discuss this in the context of the future EST that aims at correcting the turbulence within $60''\times 60''$. 
The main difficulty lies in the profile reconstruction. Indeed, the retrieval of the turbulence volume from measurements along a few discrete directions is an ill-conditioned problem: 
with a finite number of sensing directions, a substantial fraction of the atmospheric volume is sensed by only one sensor (see Fig.\,\ref{fig:triang}). Triangulation is then not possible and the origin of the distortion can not be determined. Even when two sensors measure a correlated signal, the correlation might accidentally originate from two different turbulent cells. 
The control loop of a star-oriented MCAO system is therefore fed with prior information about the Kolmogorov nature of atmospheric turbulence and about the current atmospheric profile\,\cite{Neichel},\,\cite{Rigaut2010}. The indetermination in the tomographic reconstruction is then not solved, but the  correction is -- on average -- more right than wrong\,\cite{Rigaut2000}.

\subsection{Layer-oriented MCAO} 

\begin{figure}[htbp]
\begin{center}
\includegraphics[width=.6\textwidth]{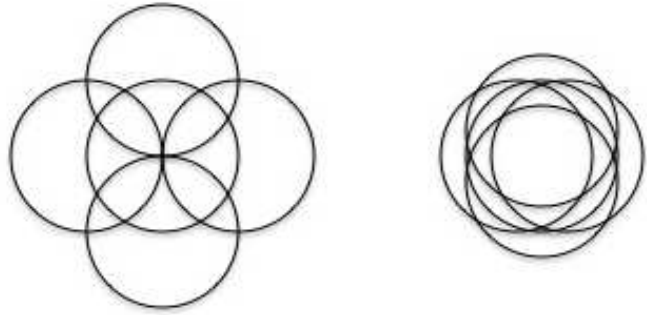}
\caption{ Principle of the layer-oriented approach: the wavefront distortions are introduced at an altitude where the wavefronts from 5 stars are disposed as shown in the left panel. If the sensor is conjugated to an altitude where the wavefronts are disposed as shown in the right panel, the distortions are smoothed out and the sensor signal is attenuated.}
\label{fig:LO}
\end{center}
\end{figure}

The  layer-oriented approach to MCAO has been introduced for nighttime observations\,\cite{Ragazzoni}, \cite{Ragazzoni2}, \cite{Arcid}:
each deformable mirror works in closed loop with a wavefront sensor -- mirror and sensor being conjugated to the same altitude.
The principle employs pyramid wavefront sensors. For each deformable mirror, a group of pyramid sensors images different stars onto the same location in the focal plane and the wavefront distortions are sensed on the superposed image. Distortions generated at the conjugate altitude do not vary within the field and are unaffected by the averaging, while distortions introduced at a large distance tend to cancel (see Fig.\,\ref{fig:LO}).
The signal measured by the sensor is thus an approximation of close-by turbulence. 

There are many advantages to nighttime layer-oriented MCAO, but the inherent limitation in the quality of the profile reconstruction remains unchanged: due to the finite number of reference stars, a substantial fraction of the atmospheric volume is not accessible to triangulation (see Fig.\,\ref{fig:triang}). 
The Sun however allows for an infinite number of reference targets within the field-of-view, so that the indetermination in the profile reconstruction can in principal be overcome. The use of several pyramid-sensors is then unsuitable, since we aim at averaging the wavefront distortions continuously over the field. A continuous sampling would require an infinite number of pyramid sensors -- or, at least, as many sensors as there are isoplanatic patches within the field-of-view.

\section{Solar layer-oriented MCAO} 

\subsection{Design}

We propose to implement the layer-oriented approach in solar MCAO using Shack-Hartmann (SH) sensors. 
AO systems for solar observations are already based on these sensors\,\cite{Rimmele2011}. 
Their use introduces a difficulty compared to nighttime observations on distant stars: 
when the object observed with a SH sensor is a point-source, typically a distant star, each lenslet forms a disc-like $2\times 2$ pixel image and the image-centers are computed in terms of a  photocenter calculation. 
Solar images are extended and the wavefront shifts need to be assessed by correlating $\sim20\times 20$ pixel images. The size of the image is a compromise between the quality of the correlation (a large enough image with sufficient details) and the requirement to sense the wavefront distortions within a narrow field. Indeed, the distortions are averaged over a surface that increases with altitude and field size, so that the signal from high layers gets attenuated on larger fields. 

We suggest to associate a SH sensor to each deformable mirror and to purposely correlate wide-field images.   
Distortions introduced close to the conjugate height of the mirror do not vary within the field, while distortions from distant layers vary and tend to cancel. The reconstruction of the turbulence profile is thus done optically. 

\begin{figure}[htbp]
\begin{center}
\includegraphics[width=.9\textwidth]{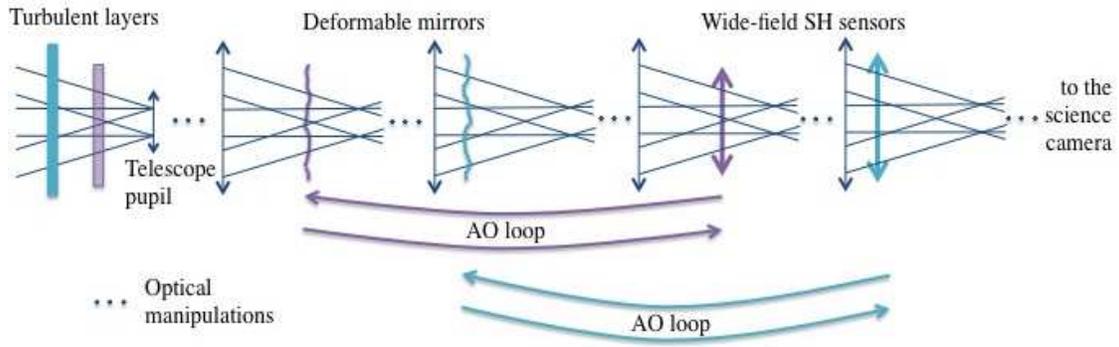}
\caption{ Principle of the layer-oriented approach for solar observations: the MCAO system consists of independent AO loops.  Each loop contains a sensor and a mirror conjugated to a dominant turbulent layer. The SH sensors measure the average wavefront distortions inside the entire field-of-view. The process of averaging attenuates signal from distant layers. }
\label{fig:layout}
\end{center}
\end{figure}

\begin{figure}[htbp]
\begin{center}
\includegraphics[width=.9\textwidth]{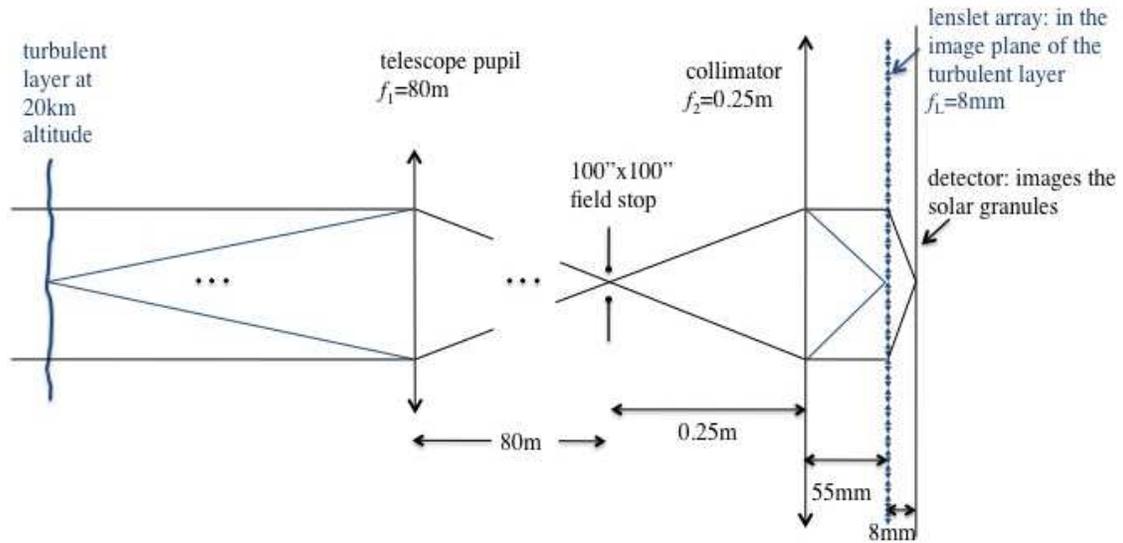}
\caption{ Design example for a 2\,m telescope with 80\,m focal length and $100''$ field-of-view. A layer at 20\,km altitude with a Fried parameter $r_0=0.4\,$m is imaged onto the SH-array. The resulting parameter values for the lenslet array and the detector are indicated in the text. }
\label{fig:layout2}
\end{center}
\end{figure}

The principle is sketched on Fig.\,\ref{fig:layout}. 
Fig.\,\ref{fig:layout2} details the design for a 2\,m diameter telescope with a $f_1=80$\,m focal length and an $\alpha=100''$ diameter field-of-view. Assume a layer at 20\,km distance with a Fried parameter $r_0=0.4$\,m. If the collimator has a focal length $f_2=0.25$\,m, a convenient diameter for the SH lenslets is: $p=r_0\,f_2/f_1=1.25\,$mm. This ensures one sampling point per Fried cell. 
The diameter of the meta-pupil at 20\,km equals 11.7\,m, the SH array should thus consist of $29\times29$ lenslets. 
The images from the SH-lenslets are adjacent without overlap, if the focal length of the SH-lenslets equals: $f_L=r_0\,(f_2/f_1)^2/\alpha=8\,$mm. 
The angular resolution is determined by the size of the pixels: it equals $0.''56$ ($1.''2$) for $7\,\mu$m ($15\,\mu$m) pixels. The cross-correlation is then done over $180\times 180$ pixel ($85\times 85$ pixel) images, at 1-2\,kHz. The final choice is a compromise between calculation times (not too many pixels) and the quality of the cross-correlation (enough details, hence small pixels). 

A similar set-up was proposed by R.\,Dunn in order to measure the instrumental aberrations within the field\,\cite{Dunn}. 
The possibility to use SH sensors for layer-oriented MCAO has been proposed by E. Ribak for nighttime astronomy\,\cite{Ribak}. To our knowledge it has not yet been put into practice. The main attraction of the method proposed by R.\,Ragazzoni is the co-adding of light from several stars. Faint stars that can not be used as reference when imaged alone, can then contribute to the signal. This advantage is lost with SH sensors. 
In solar astronomy flux is not an issue, and the main advantage of the layer-oriented approach is different: 
since the entire field is used for wavefront sensing, the control-loop automatically finds the optimal correction for the entire field. In the star-oriented approach the optimal correction for the entire field is extrapolated from measurements along a finite number of directions. And the quality of that extrapolation is limited by the quality of the tomographic reconstruction. 

The sensors of a layer-oriented MCAO system should be directed towards granular patterns rather than solar spots, so that the different regions in the field contribute similarly to the signal. In the presence of bright spots the system resembles a nighttime layer-oriented MCAO. 

The main advantage of this method is the use of the entire field for wavefront sensing. Another advantage lies in the fact that the mirror-sensor pairs form independent correction loops. Each sensor can therefore be tuned to the characteristic scales of its associated layer: the sub-aperture size can be enlarged and the correction frequency decreased. The ground layer, for example, is expected to be strong but slow, so that the sensor should be designed with enough sub-apertures, but the correction frequency can be chosen below the usual 2000\,Hz. 
The advantages and drawbacks of the star- and layer-oriented approaches are summarized in Table\,\ref{tab:adv}.

\subsection{Attenuation of the signal from distant layers}\label{sec:expression}

For multi-conjugated AO the atmosphere is approximated by a finite number of layers at selected altitudes. In the layer-oriented approach to MCAO, the phase distortion due to a particular layer is measured by a sensor that is positioned in its image plane. The sensors provide a focused image of this layer, but the image contains also the out-of-focus images of the fluctuations in the other layers. These unwanted contributions are images of the fluctuation patterns that are each averaged out over circular domains that depend on the field-of-view, $\alpha$ (typically $50''-200''$) and on the altitude difference between the conjugated and the non-conjugated layer. 

As seen in the image plane of layer $i$ at altitude $h_i$ the image of layer $k$ at altitude $h_k$, is averaged over diameter, $d_{i,k}$:
\begin{eqnarray}
d_{i,k} =\alpha |h_i - h_k|	
\end{eqnarray}
At each point the phase value, $f$, is replaced by the phase average over the disc of diameter $d_{i,k}$ centered at the point. Since only the phase differences are of concern the entire phase screen is normalized to zero mean value. 

\begin{figure}[htbp]
\begin{center}
\includegraphics[height=.27\textwidth]{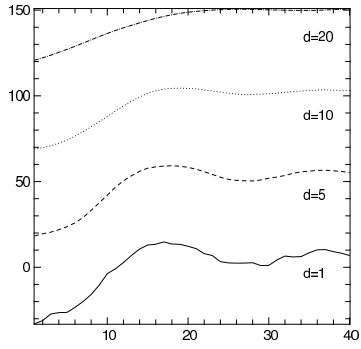}
\caption{Cross section through part of a wavefront averaged over discs of different diameter, $d$. The unit of length is taken to be the Fried parameter.  The length of the segment is 40, the points are plotted a unit distance apart. The averaged values depend on the phases within a distance up to $d/2$ from the line segment. }
\label{fig:1}
\end{center}
\end{figure}

The averaging tends to reduce the amplitude of the phase fluctuations. Fig.\,\ref{fig:1} indicates in terms of values from a simulated phase screen the character of this reduction. The lower series of phase shifts corresponds to the values measured by a linear array of 40 sensor elements positioned unit length apart; the Fried parameter is taken as unit of length; absolute values are not specified since they depend on the fraction of the atmosphere that is ascribed to a single layer. The upper curves represent the changed signals that result when the phase field is averaged over larger diameters. Although, for this example, the values relate all to the same cross-section of the same simulated phase field, the values for larger $d$ can not be deduced from those for smaller $d$, because they depend on the phase distribution in an increasingly larger domain around the line segment. As one would expect, the averaging reduces most strongly the short wave fluctuations.

To quantify the contribution of the out-of-focus image components to the entire signal one needs to compare the phase variance in the original and the degraded images. 
The variance of the wavefront phase over a circle of diameter $d$ is given by (see for example Roddier\,\cite{Roddier}):
\begin{eqnarray}
\sigma^2 (d) \propto \int_{\theta=0}^{2\pi} \int_{\nu=0}^{+\infty} W_F (\vec \nu) \, G_d (\vec \nu)\,{\rm d\/}\nu\,\nu\,{\rm d\/}\theta 
\end{eqnarray}
where $W_F$ is the power spectrum of the phase fluctuations and $G_d$ is the point-spread function through a circular opening of diameter $d$:
\begin{eqnarray}
W_F (\vec \nu) &\propto& \left(\nu^2+\frac{1}{L_0^2}\right)^{-11/6} \\
G_d (\vec \nu) &\propto& \left(\frac{{\rm J_1}(\pi \nu d)}{\pi \nu d}\right)^2
\end{eqnarray}
$L_0$ is the outer scale of turbulence. In the case of square sub-apertures, typically used with SH sensors, $G_d (\vec \nu) \propto {\rm sinc\/}^2(\pi \nu d)$. Circular openings are assumed in the following, but the calculations can easily be repeated for square apertures. 

\begin{figure}[htbp]
\begin{center}
\includegraphics[width=6cm]{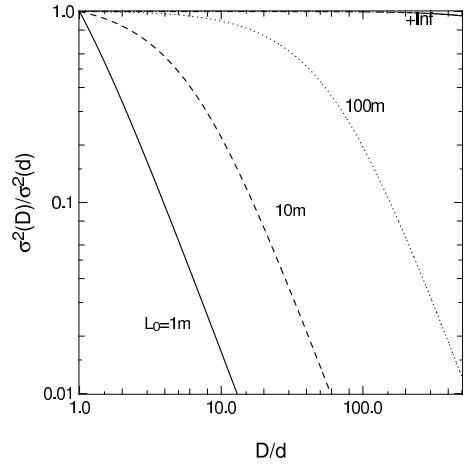}
\caption{Variance of the mean wavefront phase over circles of diameter $D$ and $d$ (see Eq.\,\ref{eq:var_deg}). For a small outer scale, $L_0$, the attenuation of the signal from distant layers is more efficient. }
\label{fig:blurring}
\end{center}
\end{figure}

An integration over all directions of the frequency plane leads to: 
\begin{eqnarray}
\sigma^2 (d) \propto \int_{\nu=0}^{+\infty}  \nu\, \left(\nu^2+\frac{1}{L_0^2}\right)^{-11/6} \, \left(\frac{{\rm J_1}(\pi \nu d)}{\pi \nu d}\right)^2\,{\rm d\/}\nu
\label{eq:var_deg}
\end{eqnarray}
Fig.\,\ref{fig:blurring} illustrates $\sigma^2 (D)/ \sigma^2 (d)$ for different values of the outer scale $L_0$: the attenuation of the signal strongly depends on $L_0$. Nighttime values of the outer scale lie between a few tens and a few hundred meters\,\cite{GSM}. Measurements of daytime values are sparse, but appear to suggest much smaller values between 1 and 10\,m\,\cite{Seghouani}, \cite{Sun}. A small outer scale benefits layer-oriented MCAO, since the attenuation of the signal from distant layers is then stronger. 

\subsection{Note on phase, slope and curvature measurements}\label{sec:phslcv}

Eq.\,\ref{eq:var_deg} can be extended to the case of slope ($sl$) and curvature ($cv$) measurements: 

\begin{eqnarray}
\sigma_{sl}^2 (d) &\propto& \int_{\nu=0}^{+\infty} {\rm d\/}\nu \cdot \nu^3\,\left(\nu^2+\frac{1}{L_0^2}\right)^{-11/6} \, \left(\frac{{\rm J_1}(\pi \nu d)}{\pi \nu d}\right)^2 \label{eq:sl}\\
\sigma_{cv}^2 (d) &\propto& \int_{\nu=0}^{+\infty} {\rm d\/}\nu \cdot \nu^5\,\left(\nu^2+\frac{1}{L_0^2}\right)^{-11/6} \, \left(\frac{{\rm J_1}(\pi \nu d)}{\pi \nu d}\right)^2
\label{eq:cv}
\end{eqnarray}

\begin{figure}[htbp]
\begin{center}
\includegraphics[width=6cm]{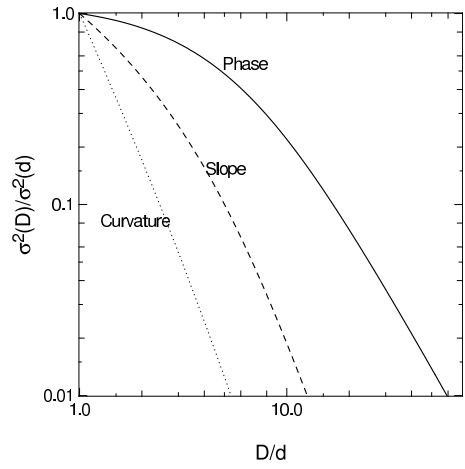}
\caption{ Relative contribution of each layer to phase, slope and curvature measurements. The curves are calculated in terms of Eqs.\,\ref{eq:var_deg}--\ref{eq:cv} with an outer scale $L_0=10$\,m. The contribution of a distant layer ($D>>d$) is smallest for curvature measurements. }
\label{fig:phslcv}
\end{center}
\end{figure}

The attenuation of the signal from a mis-conjugated layer is smallest in the case of direct phase measurements and it is largest for curvature measurements, see Fig.\,\ref{fig:phslcv}. 
It is tempting, but incorrect, to conclude that Roddier curvature-sensors are optimally fitted for layer-oriented MCAO\,\cite{Roddier-cvsensor}: no matter which quantity is measured -- phase, slopes or curvature -- the phase values need to be restored to shape the deformable mirror. This is also the case for the bi-morph mirrors that are typically  used in combination with curvature sensors. Bi-morph mirrors are controlled in curvature (the applied voltage changes the local curvature of the mirror surface), so that the phase values need not be computed by the control loop, but they are restored by the mirror itself. As indicated by F. Roddier, bi-morph mirrors solve the Poisson equation themselves\,\cite{Roddier-cvsensor}. 
Accordingly, it is indeed the attenuation of the phase values that matter in the present study.

\section{Applications}

\subsection{Ground layer adaptive optics}

\begin{figure}[htbp]
\begin{center}
\includegraphics[width=6cm]{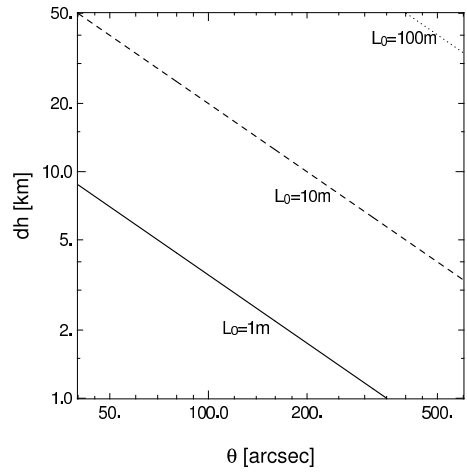}
\caption{Minimum altitude-difference between the layer and the sensor for the variance of the signal to fall below the fitting error. The $x$-axis indicates the angular diameter of the circular field-of-view. The results are derived from Fig.\,\ref{fig:blurring}. }
\label{fig:dh}
\end{center}
\end{figure}

Rimmele et al. tested a ground-layer adaptive optical (GLAO) correction at the DST in New Mexico\,\cite{Rimmele2010}: the group used a SH sensor conjugated to the ground and averaged the wavefront distortions over a $42''\times 42''$ field of view. The average slopes were used to control a deformable mirror conjugated to the ground. This should have led to the suppression of the ground layer only and thus to a homogeneous correction throughout the field. The experiment was however unsuccessful. 
Fig.\,\ref{fig:dh} shows that a SH sensor conjugated to the ground, with a $42''\times 42''$ field-of-view is sensitive to turbulence up to almost 10\,km for an outer scale of 1\,m. For a more realistic outer scale of 10\,m, the sensor is sensitive to turbulence up to 50\,km. A 5' field-diameter is required to efficiently attenuate turbulence above 5\,kilometers. 

We use the residual variance of the fitting error as a norm to assess the attenuation of the signal. The fitting error corresponds to the part of the wavefront that is not corrected by the mirror due to the finite number of actuators. 
The uncorrected phase-variance over a circular area of diameter $d$ equals\,\cite{Noll76}:
\begin{equation}
\sigma^2 =1.03\, \left( \frac{d}{r_0}\right)^{5/3}
\end{equation}

The residual phase-variance after AO correction is given by\,\cite{Sechaud99}:
\begin{eqnarray}
\label{eq:fittingerror}
\sigma_r^2 = \mu \, \left( \frac{d}{r_0}\right)^{5/3} 
\end{eqnarray}
where $d$ is the spacing between two actuators. In a layer-oriented approach, the diameter of the sub-apertures will be set equal to the actuator spacing, $d$. The value of $\mu$ depends on the shape of the actuators' {\it influence-functions\/} and equals 0.2 for typical mirrors with Gaussian like influence-functions. 
The fitting error thus amounts to 20\% of the initial phase variance over the sub-aperture. 
We assume that the signal of a layer is negligible if its variance is attenuated by a factor 5. This applies if the Fried parameter of the layer equals the sub-aperture size of the sensor, $d$: for a stronger layer the attenuation factor needs to be larger, while a weaker layer rapidly falls below the threshold of the fitting error. 

Fig.\,\ref{fig:dh} shows that the reduction of the phase-variance is slow: fields several arcminutes in diameter are required to attenuate the signal from layers above $2-3$\,km altitude. This appears to exclude the layer-oriented approach for GLAO systems, at least with current technology. 

\subsection{Multi-conjugate adaptive optics}\label{sec:gain}

In multi-conjugate adaptive optics smaller fields can be used:  each sensor is then somewhat  sensitive to turbulence that its associated mirror is not supposed to correct. 

Diolaiti et al. have demonstrated that the correction loop is nonetheless stable\,\cite{Diolaiti2001}. 
Each deformable mirror corrects for its own layer and for smoothed versions of the non-conjugated layers. 
In order to avoid an over-correction of one layer by several mirrors, we suggest to measure the amplitude of signals contributed by distant layers, and to use these measurements to adjust the gain of the AO loops.

Approximate the atmosphere by a number $L$ of horizontal layers $(l_i), i=1, 2, .. L$ at heights $h_i$. In actual measurements $L=3$ may suffice, but for the general considerations a larger number is considered.

Assume that sensors in the image planes conjugated to the layers $(l_i)$ measure the phase, $F_i(x,y)$, of the wavefronts that traverse the point $(x,y)$ in $(l_i)$.
$F_i(x,y)$ refers to one measurement by sensor $i$; it is the mean value over all rays that traverse $(x,y)$ within the circular field of view of diameter say $\alpha = 100''$. 

$F_i(x,y)$ is the mean phase due to the entire atmospheric traversal of the rays, i.e. it contains the contributions from all layers:
\begin{eqnarray}
F_i(x,y) &=& \sum_{l=1}^L f_{i,l} 
\label{eq:4}
\end{eqnarray}

$\sum_{l=1}^L f_{i,l}$ is the sum of the mean phases due to the individual layers. The averaging needs to be done over the intersection of the field, i.e. the viewing-cone, with the layer. The surface of the intersection depends on the relative distance between the layer and the sensor. 
Given a Kolmogorov phase screen (Kps), for  $l=i$ the term $f_{i,l}$ is a single random  value on this screen -- the wavefront phase, that can be directly applied to the deformable mirror. 
For  $l\neq i$   the term $f_{i,l}$  is a random value of the modified Kps that is obtained by blurring the Kps, i.e. by averaging the phase over the cross-section of size $r\times r$, where $r$ depends on the angle $\alpha$ and the separation between the layers $i$ and $l$.

In other words, the individual contributions $f_{i,l}$ can be obtained from the original Kps and its appropriate degradations. The variance of $f_{i,l}$ for specified blurring parameters, $r$, has been determined in Section\,\ref{sec:expression} (see Fig.\,\ref{fig:blurring}). 
There is, of course, an added coefficient $g_l=C_n^2(l)\,{\rm d\/}h$ that depends on the layer thickness and its turbulence intensity. 
 
Consider the variance $g_l\,\sigma_{i,l}^2$ of $f_{i,l}$. The fluctuations in separate layers are statistically independent. Accordingly the contributions, $\sigma_{i,l}^2$, to the variance $\sigma_i^2$ add up:
\begin{eqnarray}
\sigma_i^2 = \sum_{l=1}^L g_l\,\sigma_{i,l}^2
\end{eqnarray}

$\sigma_{i,l}^2$ is the variance of the degraded Kps for unit thickness and unit turbulence intensity of the layer. $g_l$ is the weight factor which equals the product of the layer thickness and the turbulence intensity. 

The original and the blurred phase screens provide the parameters $\sigma_{i,l}^2$ and the $\sigma_i^2$ are measured. Since there are $L$ equations, one readily computes the $L$ weight factors $g_l$, i.e. the relative contributions of the layers to the phase fluctuations.

Knowing the $g_l$ facilitates then the computation of the phase distortions $f_{i,i}(x,y)$ caused at the various locations of the different layers on the basis of the observed mean phases, $F_i(x,y)$, at these locations. The $f_i(x,y)$ are the values for correcting layer $(l_i)$.

\section{Practicality of the star- and layer-oriented approaches}

Approximate the atmosphere by $L$ layers at the conjugate altitudes, $h_i$, of the deformable mirrors. 
Each layer contains a fraction $f_i$ of the turbulent energy and the Fried parameter in each layer equals $r_0/f_i^{3/5}$. 
Let $\alpha$ be the angular diameter of the  corrected field and $D$ the telescope diameter. 
The number of actuators on each deformable mirror equals:  $\left((D+h_i\,\alpha)\, f_i^{3/5}/r_0\right)^2$, where  $D$ is the telescope diameter. 

\begin{itemize}
\item[--] In the layer-oriented approach the number of sensors equals the number of deformable mirrors, $L$, and each sensor has as many sub-apertures as there are actuators on the deformable mirror. Strictly speaking there are 4 actuators at the corners of each sub-aperture. But, as the number of actuators increases, the number of sub-apertures tends towards the number of actuators. The total number of sub-apertures, $N_S$, then equals:
\begin{eqnarray}
N_S = \sum_{l=1}^L \left((D+h_l\,\alpha)\, f_l^{3/5}/r_0\right)^2  \label{eq:N1}
\end{eqnarray} 

\item[--] In a star-oriented approach, the sensing stage typically consists of one high-order on axis sensor with $(D/r_0)^2$ sub-apertures, and one low order, wide-field sensor with $(D/r_H)^2$ sub-apertures. $r_H=r_0\,(\sum_{l=2}^{L} f_l)^{-3/5}$ is the high-altitude Fried parameter. 
The number of sub-apertures should be larger than the number of actuators:
\begin{eqnarray}
N_S&=&\left(\frac{D}{r_0}\right)^2 +(S-1)\,\left(\frac{D}{r_H}\right)^2  \geq  \sum_{l=1}^{L} \left(\frac{(D+h_l\, \alpha)\, f_l^{3/5}}{r_0} \right)^2 \nonumber \\
S &\geq& \frac{\sum_{l=1}^L (1+h_l\,\alpha/D)^2\,f_l^{6/5} -1}{\sum_{l=2}^L f_l^{6/5}} +1  \label{eq:N2}
\end{eqnarray}
$S$:  number of sensing directions. 

In addition, the meta-pupils should be correctly covered. If the coverage is incomplete, some actuator voltages need to be extrapolated. Since the mirror with the highest conjugation-altitude, $h_L$, has the largest meta-pupil, this requirement translates into:
\begin{eqnarray}
S \,D^2 \geq (D+h_L\,\alpha)^2 \nonumber \\
S \geq (1+h_L\,\alpha/D)^2 \label{eq:N3}
\end{eqnarray}
\end{itemize}

\begin{figure}[htbp]
\begin{center}
\includegraphics[width=.3\textwidth]{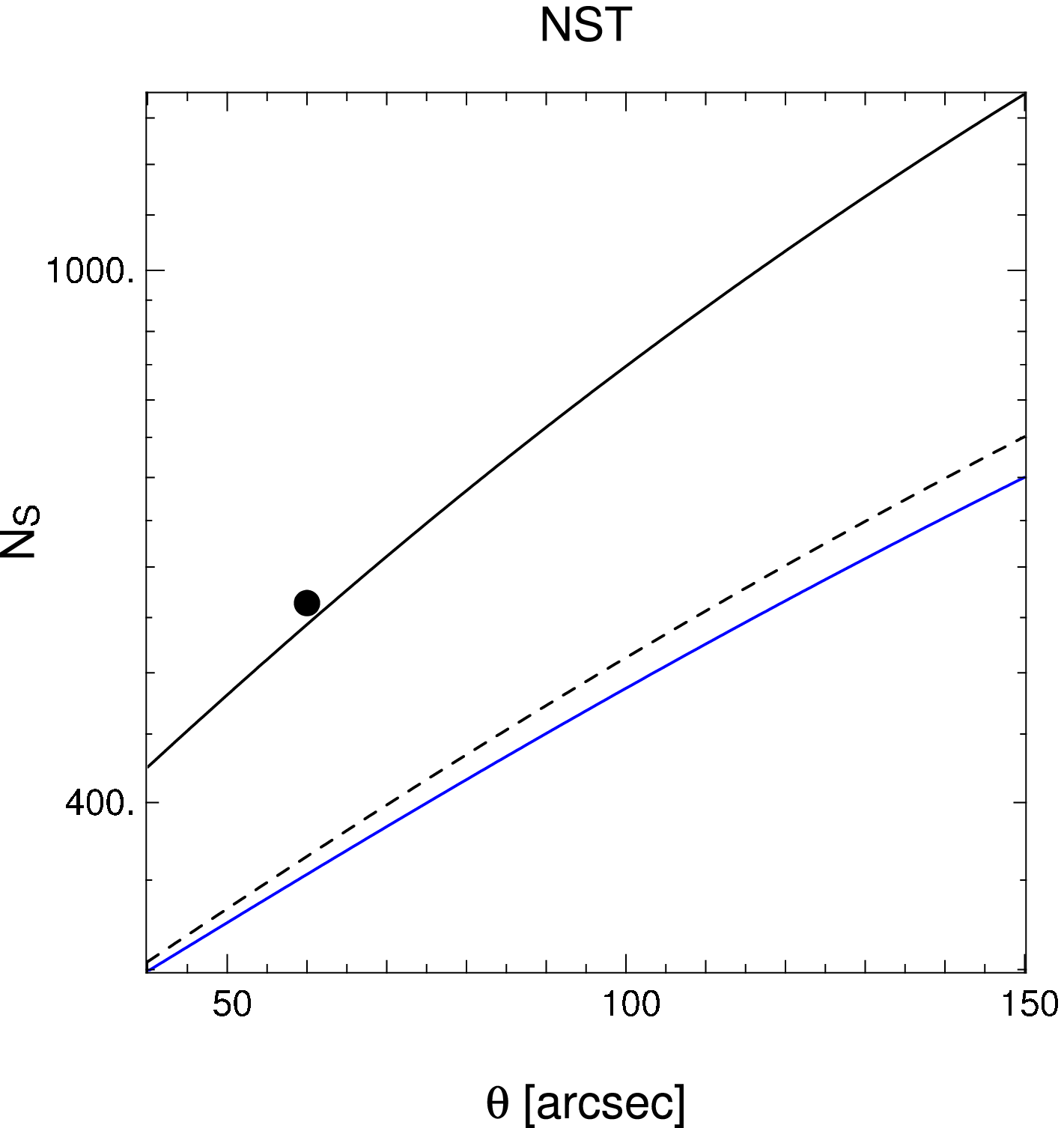}
\includegraphics[width=.3\textwidth]{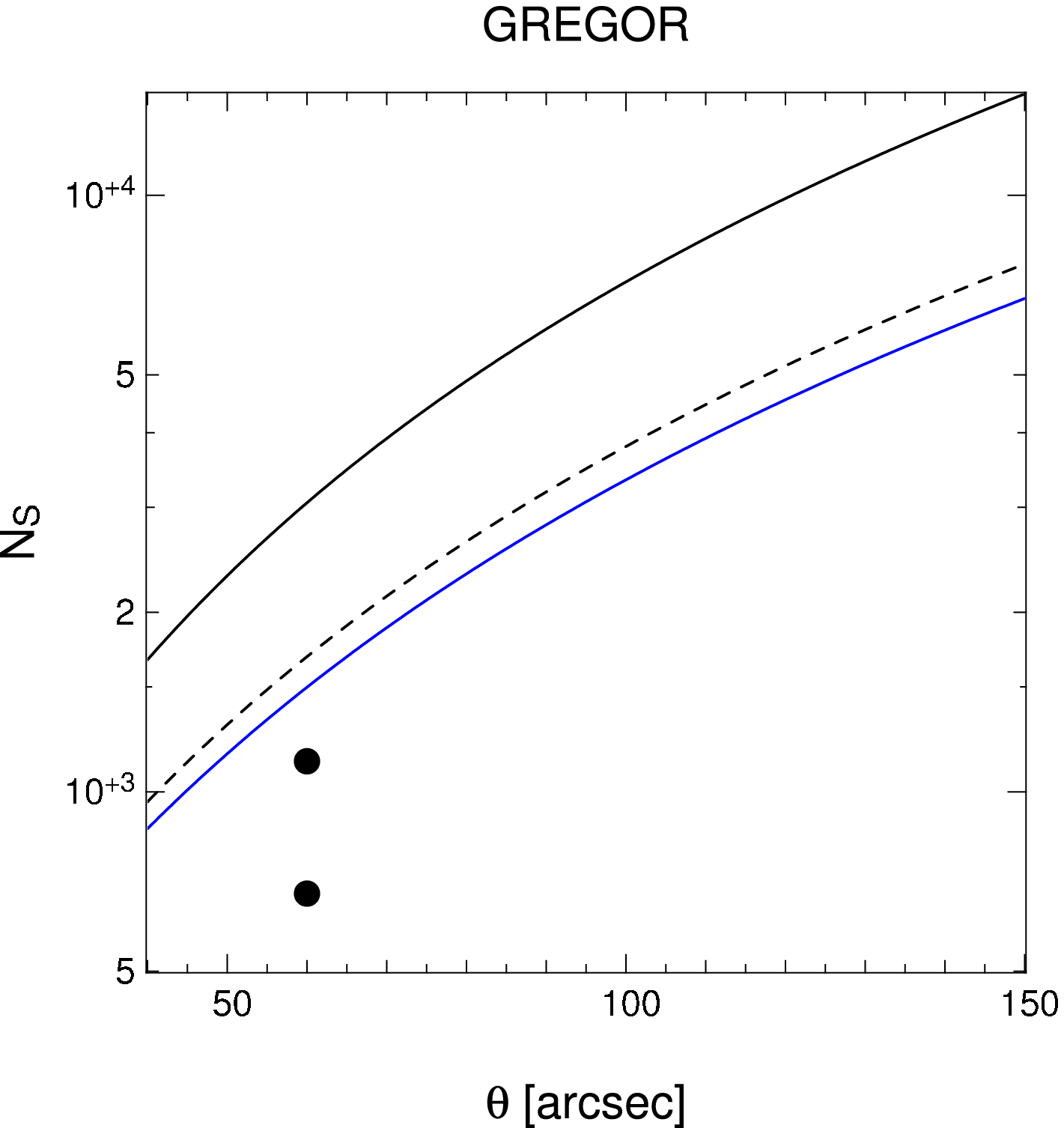}
\includegraphics[width=.3\textwidth]{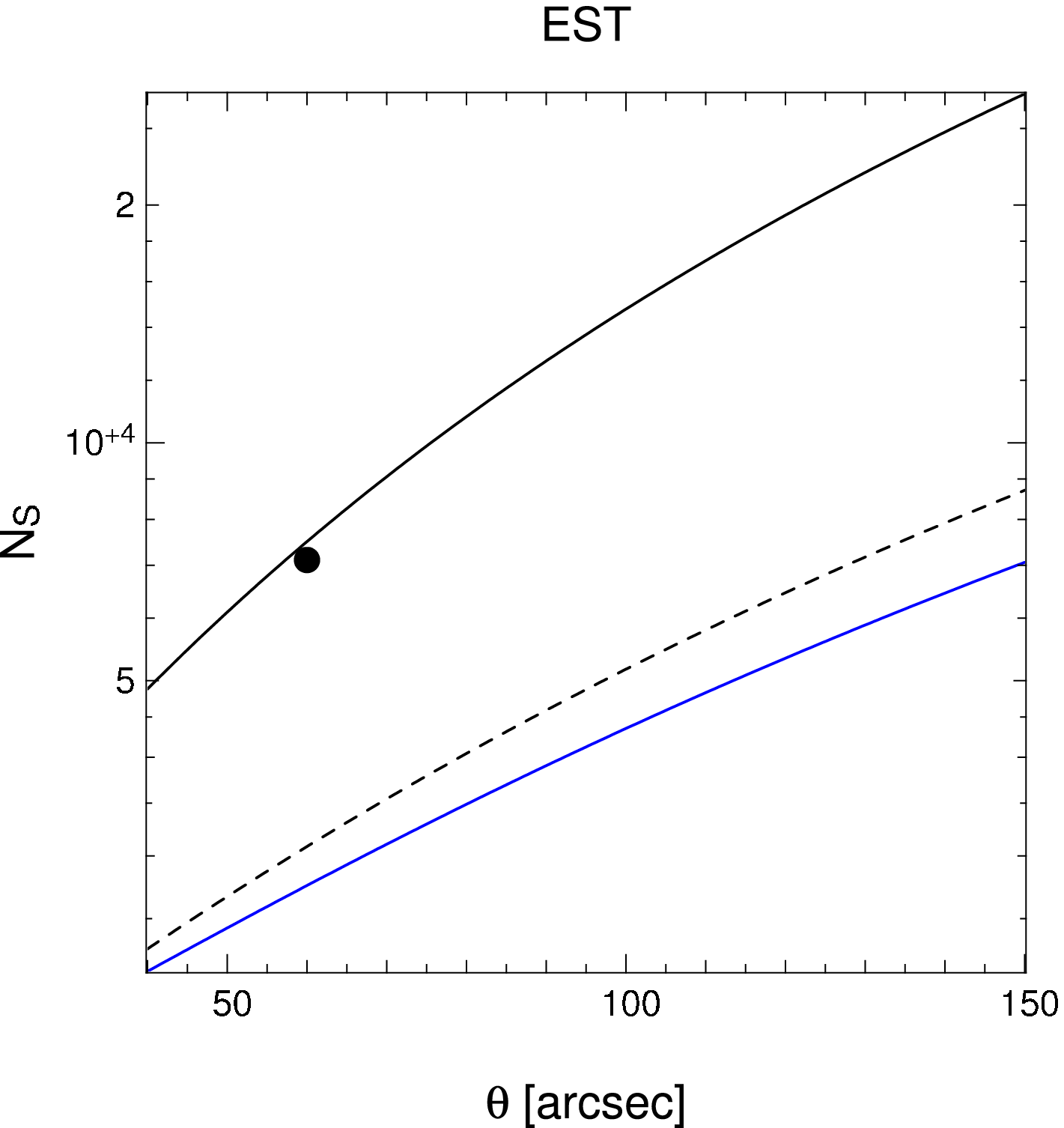}
\caption{Number of sub-apertures required in the layer and star-oriented approaches (resp. blue and black). Full black line:  the meta-pupil of the highest deformable mirror is entirely sensed, dashed black line: as many sub-apertures as actuators. Circles:  number of sub-apertures for the MCAO systems on NST, GREGOR and EST. The characteristics  of these systems are listed, with references, in Table\,\ref{tab:subap}. }
\label{fig:subap}
\end{center}
\end{figure}

The resulting numbers of sub-apertures are represented on Fig.\,\ref{fig:subap}. 
The numbers of sub-apertures obtained from Eqs.\,\ref{eq:N1} and \ref{eq:N2} are similar since they reflect the same requirement:  as many sub-apertures as actuators. For the star-oriented approach, this leads to an insufficient coverage of the mirror-pupils. 
The condition of well sampled meta-pupils (Eq.\ref{eq:N3}) implies more sub-apertures, especially for the GREGOR and EST. This is because GREGOR and EST are (will be) located on mountain sites, where observations are carried out in the morning hours with large zenith angles. A given turbulent layer appears more distant and the meta-pupils are larger than at the NST. 

\begin{figure}[htbp]
\begin{center}
\includegraphics[width=.3\textwidth]{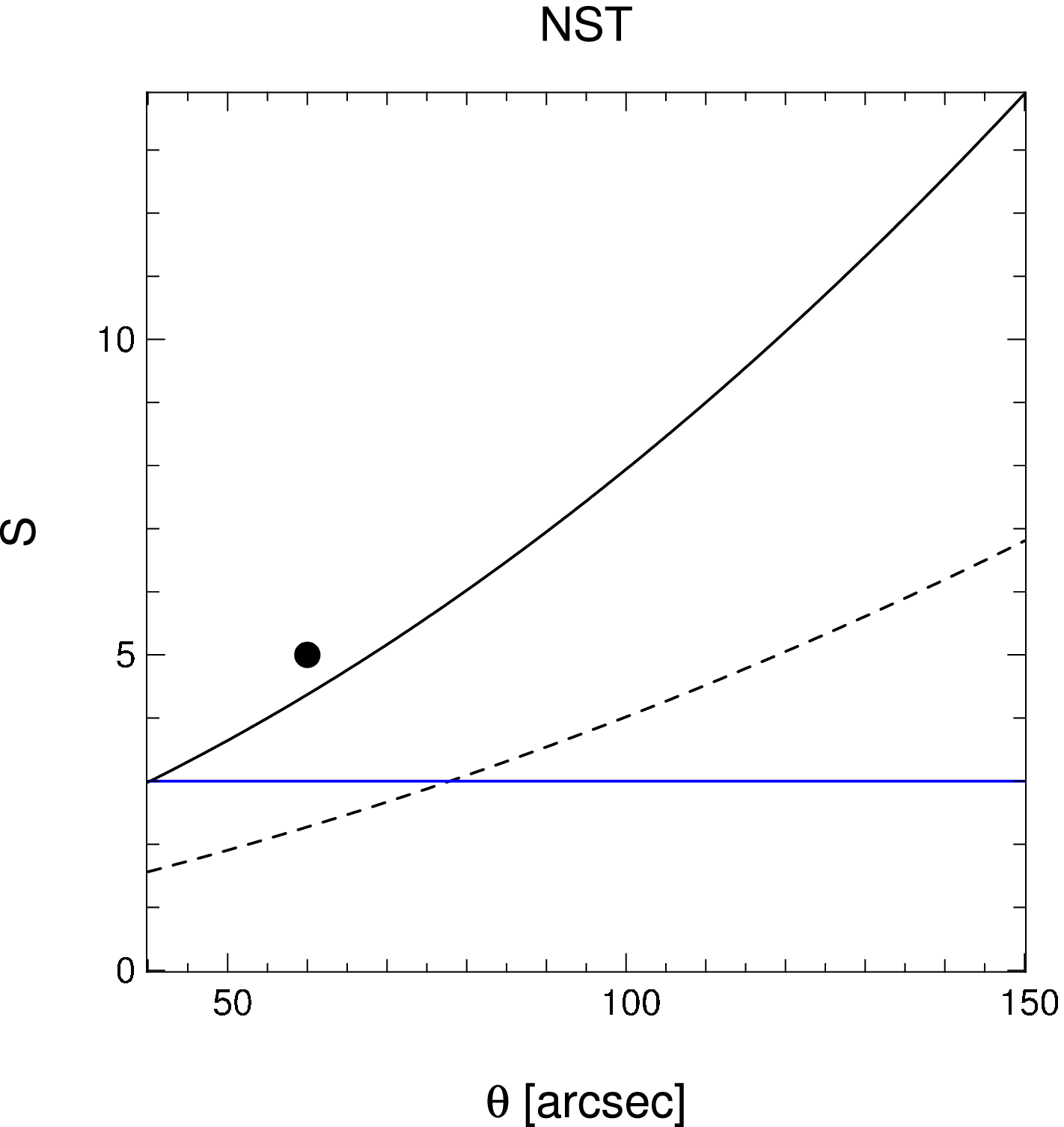}
\includegraphics[width=.3\textwidth]{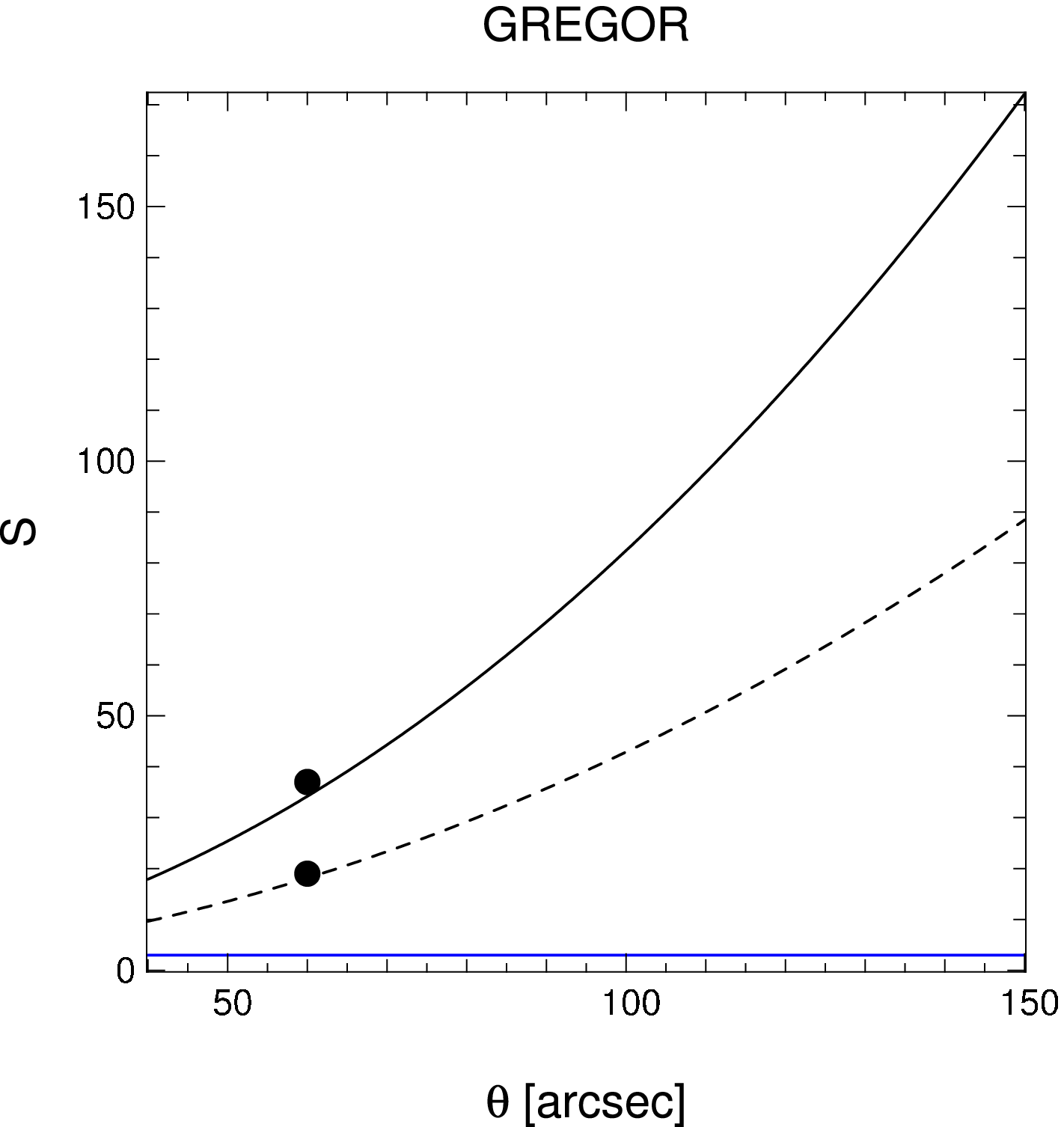}
\includegraphics[width=.3\textwidth]{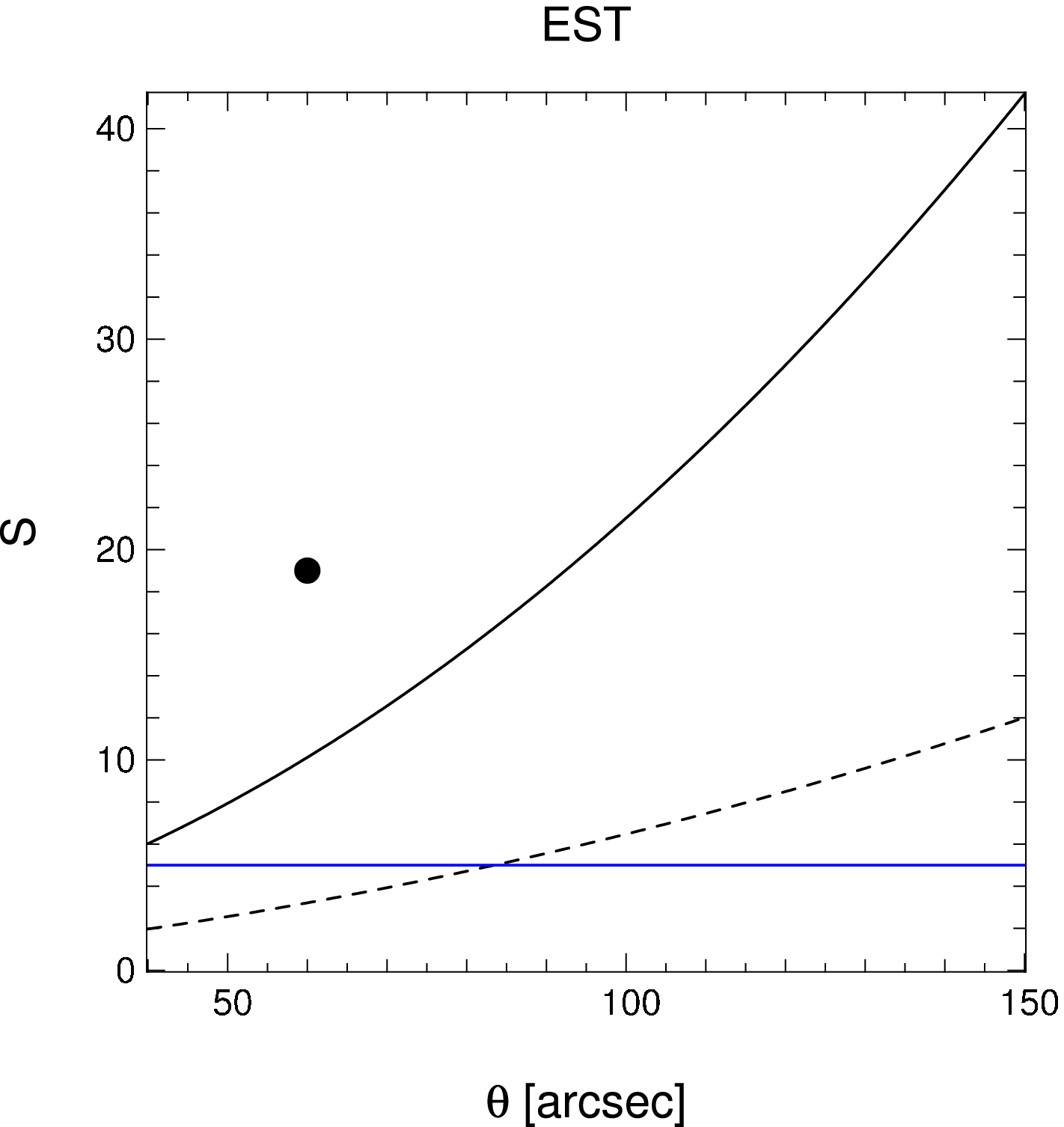}
\caption{Black lines: Number of sensing directions required in the star-oriented approach. Full line:  the meta-pupil of the highest deformable mirror is entirely sensed, dashed line: as many sub-apertures as actuators. Circles: number of sensing directions for the MCAO systems on NST, GREGOR and EST. Blue line: Number of sensors required in the layer-oriented approach. The characteristics  of these systems are listed, with references, in Table\,\ref{tab:subap}. }
\label{fig:directions}
\end{center}
\end{figure}

As the field-size increases, the number of sensing directions in the star-oriented approach becomes prohibitively large, see Fig.\,\ref{fig:directions}. 
One will then eventually opt for configurations where the mirror-pupils are incompletely sensed. 
This is already the case with GREGOR: the 19 sensing directions cover a 33\,m$^2$ surface at the highest conjugate altitude of 25\,km -- well below the 60\,m$^2$ surface of the mirror pupil within the $60''\times 60''$ field of view. 
The planned upgrade to 37 sensing directions will permit a complete coverage (65\,m$^2$) of the highest meta-pupil. 

In the layer-oriented approach the slopes are averaged over the field-of-view and all meta-pupils are entirely sensed. 
It is interesting to note the complementarity of the layer- and star-oriented approaches: 
the difficulty involved with the tomographic reconstruction appears to limit the possibility of the star-oriented approach to correct field-sizes beyond roughly $60''$, see for example the discussion in Berkefeld\,\cite{Berkefeld}. The layer-oriented approach  
fails below $\sim50''$ and benefits from increasingly large fields-of-view. 

\section{Conclusion}

We have described a layer-oriented approach to solar multi-conjugate adaptive optics. The implementation is based on cross-correlating Shack-Hartmann sensors that are already widely used for solar AO systems. Each deformable mirror should be paired with a wide-field Shack-Hartmann sensor: the sensor and the mirror are conjugated to the same altitude and work in a closed loop. The sensor measures the average wavefront distortion over the entire field-of-view via a cross-correlation of a wide-field image. The process of averaging attenuates the signal from distant layers and the sensor signal represents then adequately the nearby turbulence. The tomographic reconstruction is done optically. 

The main advantage of the approach is that the wavefront distortions are sensed within the entire field-of-view. The quality of the profile reconstruction is thus enhanced with respect to star-oriented MCAO where the turbulence is sensed along a few discrete directions. In addition, each mirror-sensor pair forms an independent control loop, the parameters of which merely need to account for local turbulence: the sub-apertures can be enlarged and the correction frequency reduced. 

We have derived the altitude-sensitivity of a sensor as a function of field-size and outer scale: the attenuation of the signal is slow and appears to exclude the use of the layer-oriented approach for ground layer adaptive optics. In a multi-conjugate system, each mirror corrects its conjugate layer and smoothed images of the mis-conjugated layers. The AO correction is stable as long as the loop-gains are not too high. A procedure to adjust the gains has been suggested. 

The layer-oriented approach benefits from larger field-sizes. In contrast, the currently used star-oriented approach is an extension of a conventional AO system and becomes more complex as the field size increases -- in terms of required number of sub-apertures and computational load.  
The layer-oriented approach will thus be an attractive solution for the future generation of solar MCAO systems.

\section*{Acknowledgements}
I am grateful to G\"oran Scharmer for valuable suggestions. Thanks to Nicolas Gorceix for helping with the design example. 
The  National Science Foundation is acknowledged for funding this research through grant NSF-AST-0079482.

\clearpage

\clearpage

\clearpage

 \clearpage

 \clearpage
 
 \clearpage

\clearpage

\clearpage

\clearpage

\clearpage

\clearpage

\begin{table*}
\centering
\begin{tabular}{|p{7cm}|p{7cm}|}  \hline
{\bf Star-oriented\/} & {\bf Layer-oriented\/}\\ \hline
 \textcolor{blue}{Only 2 detectors} &  \textcolor{red}{As many detectors as deformable mirrors}\\\hline
 \textcolor{red}{Distortions not sensed over the entire field-of-view} &  \textcolor{blue}{Distortions sensed over the entire field-of-view} \\\hline
 \textcolor{red}{AO loop frequency imposed by the fastest layer, typically $>2000$Hz. Sub-aperture size imposed by the strongest layer}
&  \textcolor{blue}{Each sensor is tuned to the characteristic scales -- $r_0(h)$ and $\tau_0(h)$ -- of its layer: larger sub-apertures, lower correction frequencies} \\\hline
 \textcolor{red}{Correlation on few pixels to minimize field extension: larger noise on slope estimates} &  \textcolor{blue}{Correlation on many pixels: less noise on slope estimates}\\ \hline
 \textcolor{blue}{Correlation on few pixels: faster computation} &  \textcolor{red}{Cross-correlation on many pixels: longer computation times}\\
\hline
 \end{tabular}
\caption{ Qualitative comparison of the star- and layer-oriented approaches for solar MCAO. Advantages in blue, drawbacks in red. }
\label{tab:adv}
\centering
 \end{table*} 
 \clearpage
 
\begin{table*}
\centering
\begin{tabular}{| l p{1.8cm} p{4.5cm} p{2cm} p{4cm} |}  \hline
& Telescope diameter & Mirror altitudes and associated $r_0$ & Field directions & Sub-aperture size -- on/off axis sensors \\ \hline
NST & 1.6\,m&0 -- 3 -- 6 km& 5& 8-- 25 cm\\
 &&14 -- 21 -- 40 cm &&\\\hline
GREGOR & 1.5\,m& 0 -- 8 -- 25 km& 19, 37&10 -- 30 cm\\
 &&14 -- 23 -- 26 cm&&\\\hline
EST & 4\,m& 0 -- 5 -- 9 -- 15 -- 30 km& 19& 8 -- 25 cm\\
 &&14-- 30 -- 41 -- 33 -- 48 cm&&\\
\hline
\end{tabular}
\caption{ Parameter values of three planned MCAO systems: NST in California, GREGOR and EST on Canary Islands. The parameter values for GREGOR and EST are taken from Berkefeld et al.\cite{Berkefeld}, \cite{BerkefeldAO}. The values for NST were determined from profiles of the atmospheric turbulence \cite{SDIMM}. All three systems use the star-oriented approach inside a $60''\times 60''$ field. This table is used for Figs.\,\ref{fig:subap} and \ref{fig:directions}.}
\label{tab:subap}
\centering
 \end{table*}

\end{document}